\documentclass[doublecol]{epl2} 

\title{Molecular kinetic analysis of a finite-time Carnot cycle}
\shorttitle{Molecular kinetic analysis of a finite-time Carnot cycle} 

\author{Y. Izumida\thanks{E-mail:
\email{izumida@statphys.sci.hokudai.ac.jp}} 
and K. Okuda\thanks{E-mail: \email{okuda@statphys.sci.hokudai.ac.jp}}}
\shortauthor{Y. Izumida and K. Okuda}

\institute{                    
Division of Physics, Hokkaido University, Sapporo 060-0810, Japan
}

\abstract{We study the efficiency at the maximal power $\eta_\mathrm{max}$ 
of a finite-time Carnot cycle of a weakly interacting gas
which we can regard as a nearly ideal gas.
In several systems interacting with the hot and cold reservoirs
of the temperatures $T_\mathrm{h}$ and $T_\mathrm{c}$, respectively, it
is known that $\eta_\mathrm{max}=1-\sqrt{{T_\mathrm{c}}/{T_\mathrm{h}}}$ which is
often called the Curzon-Ahlborn (CA) efficiency $\eta_\mathrm{CA}$. 
For the first time numerical experiments to verify the validity of
$\eta_{\mathrm{CA}}$ are performed by means of molecular 
dynamics simulations and reveal that 
our $\eta_\mathrm{max}$ does not always agree with $\eta_\mathrm{CA}$, but 
approaches $\eta_\mathrm{CA}$ in the limit of 
$T_\mathrm{c} \rightarrow T_\mathrm{h}$. 
Our molecular kinetic analysis explains the above facts theoretically by  
using only elementary arithmetic.
}
\begin{document}

\maketitle
\section{Introduction}
Recently global warming has been a worldwide problem.
Developing more efficient engines may help to solve such a problem.
In physics, the efficiency of heat engines has been treated as a basic 
subject of thermodynamics.
One of the most important results is the discovery of the Carnot efficiency 
which gives the upper limit of efficiency: 
$\eta_\mathrm{C}=1-T_\mathrm{c}/T_\mathrm{h}$, where $T_\mathrm{h}$ and 
$T_\mathrm{c}$ are the temperatures of the hot and cold heat 
reservoirs, respectively.
In spite of the high efficiency, $\eta_\mathrm{C}$ is usually realized only
in the quasistatic limit.
This means that the Carnot heat engine is useless as a real engine 
because the power defined as output work per unit time is 0.
Real engines should work for a finite time and produce a
finite power.
Therefore, the finite-time extension of the quasistatic heat engines 
is an important subject of thermodynamics.
Curzon and Ahlborn~\cite{CA,C} (see also~\cite{N}) considered such an extension 
of the Carnot cycle and derived a simple and beautiful result:
the efficiency at the maximal power output is given by
\begin{equation}
\eta_{\mathrm{CA}}=1-\sqrt{\frac{T_\mathrm{c}}{T_\mathrm{h}}}\quad 
\mbox{(CA efficiency)} . \label{eq.1}
\end{equation}
Several theoretical studies \cite{B,LL,VB,JC,GS,K},
ranging from the heat engine working in the linear response regime 
~\cite{VB,JC,GS} to the heat engine working by a quantum mechanism 
~\cite{K} support the
validity of Eq.~(\ref{eq.1}). 
This implies that $\eta_{\mathrm{CA}}$ has some sort of universality
independent of the model details. 

In spite of its importance, 
to our knowledge, no experiments have been 
carried out to verify the validity of Eq.~(\ref{eq.1}). 
Moreover, though in~\cite{CA} the temperature differences between 
the reservoirs and the working substance are taken as the parameters 
to maximize the power, they do not seem easily controllable. Thus, 
the CA efficiency Eq.~(\ref{eq.1}) is, in our opinion, still controversial in these respects. 

In this paper, we consider a more natural extension of the quasistatic 
Carnot cycle as a model system by using a weakly interacting gas which
we can regard as a nearly ideal gas. 
By means of molecular dynamics (MD) simulations,
numerical experiments to verify the validity of the CA efficiency are
performed for the first time.
Our model also accepts theoretical analysis by using only elementary 
arithmetic.
As shown later, we can reveal the validity and the limitation of the 
CA efficiency from that analysis.
\section{Model and simulations}
\begin{figure}[h!]
\begin{center}
\includegraphics[scale=0.7]{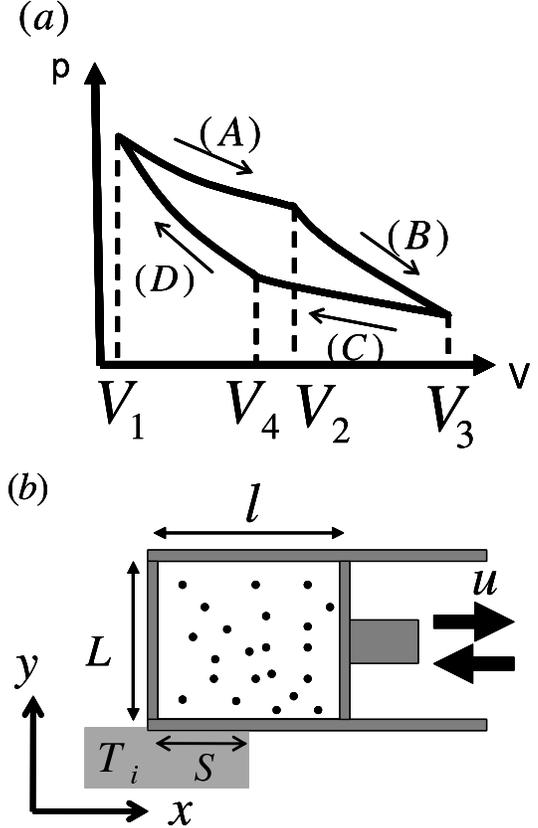}
\end{center}
\caption{Schematic illustration of the model. (a) Pressure-volume ($p$-$V$) diagram of
the quasistatic Carnot cycle for an ideal gas. (b) The piston
moves at a finite constant speed $u$  and the thermalizing wall with
the temperature $T_i$ ($i$=h, c) and the length $S$ is set on the left
bottom of the cylinder only in the isothermal processes.}\label{fig.1}
\end{figure}

We consider the quasistatic Carnot cycle of an ideal gas first and then its 
finite-time extension.
For simplicity, we here use the two-dimensional model.
The usual quasistatic Carnot cycle of an ideal gas consists of four processes:
(A): isothermal expansion process ($V_1 \to V_2$),
(B): adiabatic expansion process ($V_2 \to V_3$), 
(C): isothermal compression process ($V_3 \to V_4$),
(D): adiabatic compression process ($V_4 \to V_1$),
where $V_i$'s are the volumes of the cylinder at which we switch each of
four processes (Fig.~\ref{fig.1}(a)).
When we fix $T_\mathrm{h}, T_\mathrm{c},
 V_1$ and $V_2$, we can easily
determine the volumes $V_3$ and $V_4$ since we assume an ideal gas 
as the working substance. 
In fact, they are given by $V_3=(T_\mathrm{h}/T_\mathrm{c})V_2$ and
$V_4=(T_\mathrm{h}/T_\mathrm{c})V_1$ for the two-dimensional case.
In the case of a finite-time cycle, 
we assume that the right wall of the cylinder is a piston and moves back and forth at a constant
speed $u$.
In our model, this $u$ is taken as a unique parameter to maximize the power, 
which is controllable unlike the parameters in~\cite{CA}.
We also assume that each process is switched at the same volume 
as in the quasistatic case.

\begin{figure}
\includegraphics[width=8.5cm,height=5cm]{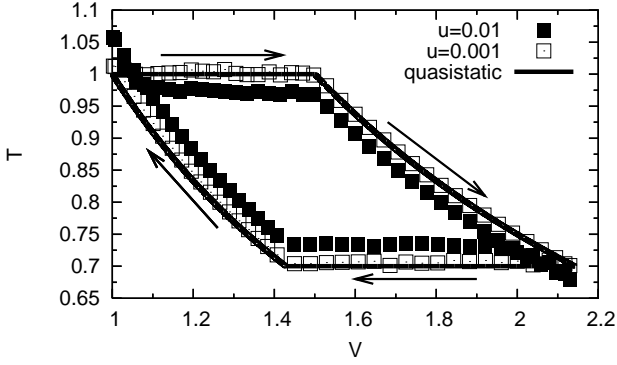}
\caption{Temperature-volume ($T$-$V$) diagram for the steady cycle at
$u=0.01$ and $u=0.001$. The data were obtained by averaging over $1000$
cycles after transient $50$ cycles in the MD simulations. The solid line is the
quasistatic Carnot cycle of an ideal gas.
We can see that as $u$ becomes larger, the cycle obviously deviates from the
quasistatic cycle.}\label{fig.2}
\end{figure}

We have performed the two-dimensional event-driven MD
simulations~\cite{AW} as follows. We assume that $N$ hard-disc particles with
diameter $d$ and mass $m$ are confined into the two-dimensional cylinder
with rectangular geometry and the collisions between hard-disc particles are perfectly elastic.
Defining $(x,y)$ coordinates as in Fig.~\ref{fig.1}(b), we let the piston move 
along the $x$-axis at a finite constant speed $u$.
Here, we express the $x$-length and the $y$-length of the
cylinder as $l$ and $L$, respectively. Then, 
the volume $V_i$ $(i=1,\cdots,4)$ of the cylinder at which we switch each of
the four processes (Fig.~\ref{fig.1}(a)) becomes $V_i=Ll_i$, where $l_i$ is
the $x$-length of the cylinder
at the switching volume $V_i$.
If the process (A) begins at time $t=0$, 
the volume $V(t)$ of the cylinder at time $t$ is given as
$V(t)=Ll(t)=L(ut+l_1)$ \ $\left(0 \le t \le
(l_3-l_1)/u \right)$
in the expansion processes (A) and (B). $V(t)$ in the
compression processes (C) and (D) is also given as
$V(t)=L(-ut+2l_3-l_1)$ \ $\left((l_3-l_1)/u \le t \le {2(l_3-l_1)}/{u}\right)$.  
When a particle with the velocity $\mbox{\boldmath$v$}=(v_x,v_y)$
collides with the piston whose $x$-velocity
is $\pm u$, its
velocity changes to $\mbox{\boldmath$v$}^{\prime}=(-v_x\pm 2u,v_y)$. Therefore, the particle
gives microscopic work
$m(|\mbox{\boldmath$v$}|^2-|\mbox{\boldmath$v$}^{\prime}|^2)/2=2m(\pm u v_x-u^2)$ against the
piston. 
In the isothermal processes, to simulate the heat reservoirs, we set the
thermalizing wall with the length $S$ at the left bottom of the cylinder
(see Fig.~\ref{fig.1}(b)).
The thermalizing wall has the following feature~\cite{YIS,H}: When a particle
collides with the thermalizing wall, its velocity stochastically changes
to the value governed by the distribution function
\begin{eqnarray}
&&f(\mbox{\boldmath$v$},T_i)=\frac{1}{\sqrt{2\pi}}\left(\frac{m}{k_\mathrm{B} T_i} \right)^{3/2}v_y \exp\left({-\frac{m\mbox{\boldmath$v$}^2}{2k_\mathrm{B} T_i}}\right),\label{eq.2}
\end{eqnarray}
($-\infty < v_x < +\infty, 0 < v_y
< +\infty$, $T_i$ ($i=\mathrm{h}$ in (A), $c$ in (C))), where
$k_\mathrm{B}$ is Boltzmann constant.
This thermalizing wall may be understood as follows.
Imagine a large particle reservoir thermalized at the temperature $T_i$
($i$=h or c) instead of the thermalizing wall and assume that if a particle 
in the cylinder goes out into the particle reservoir, another particle 
in the particle reservoir comes into the cylinder.
From this consideration, we can see that the particles coming into 
the cylinder from the particle reservoir obey the velocity distribution
function proportional to the Boltzmann factor multiplied by $v_y$.
By normalizing, we can obtain the distribution function Eq.(\ref{eq.2}).
As easily seen, this thermalizing wall guarantees that the particle 
velocities in the static system are governed by Maxwell-Boltzmann 
distribution with temperature $T_i$ :

\begin{eqnarray}
f_\mathrm{MB}(\mbox{\boldmath$v$},T_i)\equiv \frac{m}{2\pi k_\mathrm{B}T_i}\exp
\left(-\frac{m}{2k_\mathrm{B}T_i}{\mbox{\boldmath$v$}}^2\right).\label{eq.3}
\end{eqnarray}
The heat flowing from the thermalizing wall into the system can
microscopically be calculated by the difference between the kinetic 
energies before and after the collision on the thermalizing wall. 
We sum up the above microscopic heat during the simulation as
well as the microscopic work. At the walls except the piston and
the thermalizing wall, we adopt the
reflecting boundary conditions for colliding particles. We have used
$N=100$ particles with $d=0.01$ and $m=1$ in the system with $L=1, l_1=1$, $l_2=1.5$, $T_\mathrm{h}=1$,
$T_\mathrm{c}=0.7$, $k_\mathrm{B}=1$ and $S=0.5$. These
parameters except $T_\mathrm{c}$ are fixed in our all simulations and
analysis below. 
As time progresses, thermodynamic variables should draw a steady cycle
independent of initial states.
Fig.~\ref{fig.2} shows the temperature-volume diagram 
for the steady cycle at $u=0.01$ and $u=0.001$, where
$k_\mathrm{B}T$ is determined as the kinetic energy per particle,
assuming the principle of equipartition.
From this figure, we
can see that in the isothermal expansion (compression) process the temperature
approaches a steady value
lower (higher) than $T_\mathrm{h}$ $(T_\mathrm{c})$ at $u=0.01$. This result can easily be understood:
If a heat engine is working at a finite $u$, heat should flow
into the system at a finite rate to maintain the steady cycle. Therefore,
the finite difference of the temperatures between the system and the heat
reservoir is necessary. The cycle for $u=0.001$ almost agrees with the
quasistatic Carnot cycle of an ideal gas. This implies
that our system of the hard-disc particles closely approximates an ideal
gas system.      

\begin{figure}
\includegraphics[width=8cm,height=5.2cm]{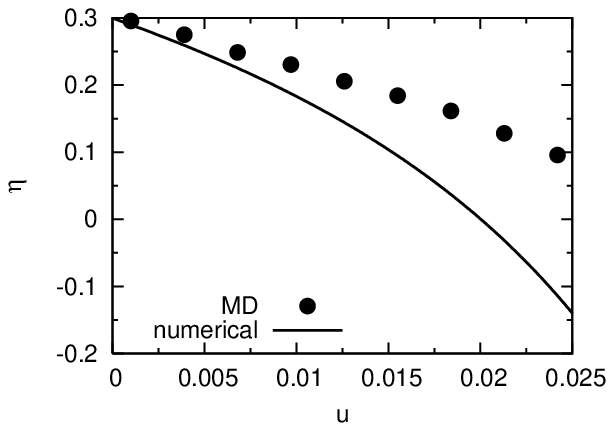}
\includegraphics[width=8cm,height=5.2cm]{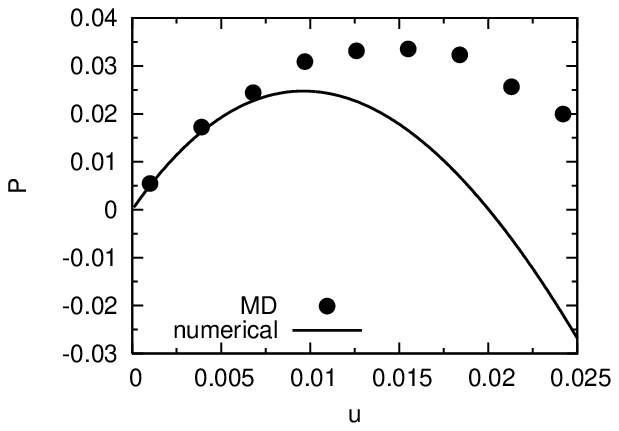}
\caption{$u$ dependence of the efficiency $\eta$ and the
power $P$. The bold dots are the values
calculated by the MD simulations and the solid line is the values
calculated by solving Eq.~(\ref{eq.9}) numerically. The MD data were
obtained by the simulations for $100$-$2000$ cycles after $35$-$50$ transient
cycles. In the limit of $u \to 0$, the Carnot efficiency ($\eta_\mathrm{C}=0.3$) is realized
and $P$ vanishes.}\label{fig.3}
\end{figure}

We have also calculated
the efficiency $\eta=W_\mathrm{total}/Q_\mathrm{h,total}$ and the
power $P=W_\mathrm{total}/{\tau}$, where
$W_\mathrm{total}$ is the total work
against the piston, $Q_\mathrm{h,total}$ is the total heat flowing
into the system from the hot heat reservoir and $\tau$ is 
the total time for the simulation.
Fig.~\ref{fig.3} shows $\eta$ and $P$ at various $u$.
We have found that the maximal power 
is realized at $u \approx 0.015$. The corresponding efficiency $\eta_{\mathrm{max}}$ (the
efficiency at the maximal power) is about 0.18, which is close to the CA
efficiency $\eta_\mathrm{CA}=0.163$.
\section{Theoretical analysis}
To explain the above MD data, we construct the theoretical model using the elementary
molecular kinetic theory as below.
We assume that even in a finite-time cycle, the gas relaxes to the
uniform equilibrium state with a well-defined temperature $T$ very fast
and the particle 
velocity $\mbox{\boldmath$v$}$ is governed by Maxwell-Boltzmann distribution
$f_\mathrm{MB}(\mbox{\boldmath$v$},T)$. We would
like to derive the time-evolution equation of $T$.
The energy $U$ of a two-dimensional equilibrium ideal gas is given by $U=Nk_\mathrm{B} T$.
In the series of cycles, $U$ can be changed by two factors: particle collisions with the
thermalizing wall and the piston. 
Our strategy to derive the time-evolution equation of $T$ is very simple: 
Counting the number of the particles colliding with the
thermalizing wall and the piston and 
calculating the heat and the work from the difference between 
the kinetic energies before and after the collisions.
Firstly, we consider the effect of the thermalizing wall.
Since the number $n_\mathrm{T}$ of the particles with the velocity
$\mbox{\boldmath$v$}$ $(v_y <0)$ colliding with the thermalizing wall
per unit time is given by
$n_\mathrm{T}=f_\mathrm{MB}(\mbox{\boldmath$v$},T)S(-v_y)N/V$, the total
number $C$ of the particles colliding with the thermalizing wall per unit time is calculated as 
\begin{eqnarray}
C=\int_{-\infty}^{+\infty}dv_x
\int_{-\infty}^{0}dv_y n_\mathrm{T}=\frac{SN}{2 \pi V}\sqrt{\frac{2 \pi
k_\mathrm{B} T}{m}}.\label{eq.4}
\end{eqnarray}
 The total energy of these colliding particles before the
collisions is also given by 
\begin{eqnarray}
\int_{-\infty}^{+\infty}dv_x
\int_{-\infty}^{0}dv_y n_\mathrm{T}\frac{m}{2}{\mbox{\boldmath$v$}}^2=\frac{3SNk_\mathrm{B} T}{4 \pi V}
\sqrt{\frac{2 \pi k_\mathrm{B} T}{m}}.\label{eq.5}
\end{eqnarray} 
Because the number of the reflecting particles is equal to the number of
the colliding particles, the total energy of the particles after the
collisions is calculated as 
\begin{eqnarray}
&&C\int_{0}^{+\infty}dv_y \int_{-\infty}^{\infty}dv_x 
f(\mbox{\boldmath$v$},T_i) \frac{m}{2}\mbox{\boldmath$v$}^2 \nonumber\\
&&=\frac{3SNk_\mathrm{B}T_i}{4\pi
V} \sqrt{\frac{2\pi k_\mathrm{B} T}{m}}\label{eq.6}
\end{eqnarray}
using Eq. (\ref{eq.2}).
Therefore, the net energy transfer, namely the heat $q_i(t,T)$ flowing
into the system per unit
time in the isothermal processes ($i=\mathrm{h}$ in (A), $\mathrm{c}$ in (C)) is given by
\begin{eqnarray}
&&q_i(t,T)=\frac{3SNk_\mathrm{B}(T_i-T)}{4\pi V(t)}\sqrt{\frac{2 \pi k_\mathrm{B}T}{m}}.\label{eq.7}
\end{eqnarray}
Next, we derive the work against the piston by the colliding
particles in the expansion processes. To calculate the number of particles colliding with the
piston, we consider the velocity distribution
$\tilde{f}_\mathrm{MB}(\tilde{v}_x,v_y,T)$ in the frame of the piston,
where $\tilde{v}_x\equiv v_x-u$ and
$\tilde{f}_\mathrm{MB}(\tilde{v}_x,v_y,T)\equiv f_\mathrm{MB}(\tilde{v}_x+u,v_y,T)$
. The number $n_\mathrm{M}$ of the particles with the
velocity $(\tilde{v}_x,v_y)$ 
colliding on the piston
per unit time is
$n_\mathrm{M}=\tilde{f}_\mathrm{MB}(\tilde{v}_x,v_y,T)L\tilde{v}_xN/V
=\tilde{f}_\mathrm{MB}(\tilde{v}_x,v_y,T)\tilde{v}_xN/l$.
Since a particle gives the work $2mu(v_x-u)=2mu\tilde{v}_x$ against the piston,
the total work $w_\mathrm{e}(t,T)$ against the piston per unit time
in the expansion processes becomes
\begin{eqnarray}
w_\mathrm{e}(t,T)
&&=\int_{-\infty}^{+\infty}dv_y
\int_{0}^{+\infty}d\tilde{v}_x 2mu\tilde{v}_x n_\mathrm{M} \nonumber\\
&&= \frac{2muN}{l(t)}
\biggl\{\frac{A^2T}{4}-Am^{3/2}\sqrt{\frac{T}{\pi}}u+\frac{u^2}{2}\nonumber \\
&&-\int_{0}^{\frac{u}{A\sqrt{T}}}d{v_x} \left(A\sqrt{T}{v_x}-u \right)^2 
\frac{\mathrm{e}^{-{v_x}^2 }}{\sqrt{\pi}}\biggr\},\label{eq.8} 
\end{eqnarray}
where $A\equiv \sqrt{{2k_\mathrm{B}}/{m}}$.
The work $w_\mathrm{c}(t,T)$ for the unit time in the compression processes is also obtained 
by changing $u \to -u$ in Eq.~(\ref{eq.8}).
  
By the energy conservation law, the time evolution of $T$ for each
of four processes (A)-(D) is given by
\begin{eqnarray}
\begin{array}{c c}
(\mathrm{A}): Nk_\mathrm{B} \displaystyle{\frac{dT}{dt}}=q_\mathrm{h}-w_\mathrm{e},&(\mathrm{B}):
Nk_\mathrm{B} \displaystyle{\frac{dT}{dt}}=-w_\mathrm{e},\\
\\
(\mathrm{C}): Nk_\mathrm{B} \displaystyle{\frac{dT}{dt}}=q_\mathrm{c}-w_\mathrm{c},&(\mathrm{D}): Nk_\mathrm{B} \displaystyle{\frac{dT}{dt}}=-w_\mathrm{c}.
\end{array}\label{eq.9}
\end{eqnarray}
Here, we have numerically solved the above Eq.~(\ref{eq.9}) for the
entire cycle.
By using the final temperature of each process as the 
initial temperature of the next process repeatedly, we can obtain 
the steady cycle of this heat engine. 
After reaching the steady
cycle, we numerically calculate the efficiency $\eta (u)={W(u)}/{Q_\mathrm{h}(u)}$ and the power
$P(u)={W(u)u}/(2(l_3-l_1))$,
where $Q_\mathrm{h}(u)$ is the heat transfer from the hot reservoir to
the system, $W(u)$ is the work output and ${2(l_3-l_1)}/{u}$ is
the time for one steady cycle.
In Fig.~\ref{fig.3}, we plot the $u$ dependence of $\eta$ and $P$ at the
same parameters as in the MD simulations. From this figure, we can see that 
the correspondence between the MD data and the line calculated
by solving Eq.~(\ref{eq.9}) numerically is established
qualitatively. This implies that our
assumption of fast relaxation to the equilibrium state is not so
bad~\footnote{Though one may see a discrepancy between the
theory and the MD simulations in Fig.~\ref{fig.3}, we can see that a
smaller $S$ gives better agreement. This is because the speed giving the
maximal power $u_\mathrm{max}$ becomes small at a small $S$, which means
that the gas is close to equilibrium and therefore meets our theoretical
assumption that the gas always stays in the equilibrium state. This
behavior of $u_\mathrm{max}$ will be confirmed in our analysis Eq.~(\ref{eq.17}).}.
\begin{figure}
\includegraphics[width=8.8cm,height=5.8cm]{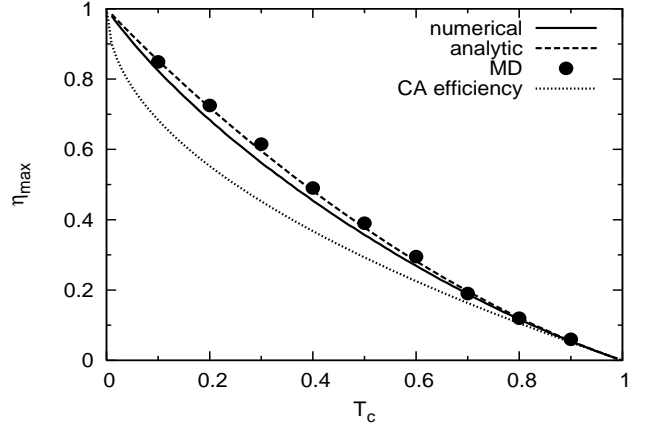}
\caption{The efficiency at the maximal power $\eta_{\mathrm{max}}$. The bold
dots are the MD data, the solid line is the theoretical
line calculated by solving Eq.~(\ref{eq.9}) numerically, the dashed line
is the the analytic expression Eq.~(\ref{eq.18}) and the dotted line is
the CA efficiency Eq.~(\ref{eq.1}).}\label{fig.4}
\end{figure}
In Fig.~\ref{fig.4}, we compare the efficiency at the maximal power
$\eta_\mathrm{max}=\eta (u_\mathrm{max})$, where $u_\mathrm{max}$ is the
speed giving the maximal power, with the CA efficiency Eq.~(\ref{eq.1}) at $T_\mathrm{h}=1$ and various $T_\mathrm{c}$. 
We have found that our
$\eta_\mathrm{max}$ does not always agree with  $\eta_\mathrm{CA}$
but tends to
approach $\eta_{\mathrm{CA}}$ as $T_\mathrm{c} \to T_\mathrm{h}$ for
both of the MD data and the numerical line. 
We have confirmed that this $\eta_\mathrm{max}$ behavior is common
to the systems with various parameters $V_1, V_2, S,$ etc., though the data are
not shown here. 
To explain this $\eta_\mathrm{max}$ behavior, we try to obtain the
analytic form of $\eta_\mathrm{max}$
by solving the evolution equation of $T$ in the following.

As seen in Fig.~\ref{fig.2}, we can expect that $T$ approaches a
steady value $T_\mathrm{h}^{\mathrm{st}}$ in the isothermal expansion
process (A). Then,
$T_\mathrm{h}^{\mathrm{st}}$ is obtained as a solution of the equation $dT/dt=0$ in
Eq.~(\ref{eq.9}A).
Because $T_\mathrm{h}^{\mathrm{st}}=T_\mathrm{h}$ is realized in the quasistatic
limit $u \to 0$, we can expand $T_\mathrm{h}^{\mathrm{st}}$ by $u$ as
$T_\mathrm{h}^{\mathrm{st}}=T_\mathrm{h}+T_\mathrm{h}^{(1)}u+T_\mathrm{h}^{(2)}u^2+{\cal
O}(u^3)$. Substituting 
$T_\mathrm{h}^{\mathrm{st}}$ into Eq.~(\ref{eq.9}A), we can determine $T_\mathrm{h}^{(1)}$ and
$T_\mathrm{h}^{(2)}$ and obtain $T_\mathrm{h}^{\mathrm{st}}$ up to
${\cal O}(u^2)$ as
\begin{eqnarray}
&&T_\mathrm{h}^{\mathrm{st}}=T_\mathrm{h}-\frac{4Lu}{3S}\biggl\{\frac{\sqrt{\pi
T_\mathrm{h}}}{A}+\frac{mu}{k_\mathrm{B}}\left(2+\frac{\pi L}{3S}\right)\biggl\}.\label{eq.10}
\end{eqnarray}
If we assume that the relaxation to $T_\mathrm{h}^{\mathrm{st}}$ is very fast, the heat
flowing into the system during $T(t)=T_\mathrm{h}^{\mathrm{st}}$ is
given by
\begin{eqnarray}
Q_\mathrm{h}^{\mathrm{st}}&&=\int_{0}^{(l_2-l_1)/u}
q_\mathrm{h}(t,T_\mathrm{h}^{\mathrm{st}})\ dt \nonumber \\
&&=Q_\mathrm{h}^{\mathrm{qs}}-2mNA\sqrt{T_\mathrm{h}\pi}\left(\frac{1}{\pi}+\frac{L}{3S}\right)u
\ln\frac{V_2}{V_1},\label{eq.11}
\end{eqnarray}
using Eq.~(\ref{eq.7}), where the quasistatic heat for ideal gas in the isothermal expansion
process is defined as $Q_\mathrm{h}^{\mathrm{qs}}\equiv Nk_\mathrm{B}
T_\mathrm{h} \ln (V_2/V_1)$. Note that $Q_\mathrm{h}^{\mathrm{st}}
\to Q_\mathrm{h}^{\mathrm{qs}}$ when we consider the
quasistatic limit $u \to 0$. 
$T_\mathrm{c}^{\mathrm{st}}$ and $Q_\mathrm{c}^{\mathrm{st}}$ of the
isothermal compression process (C) can be obtained by replacing
$T_\mathrm{h}, u, V_1$ and $V_2$ in Eqs.~(\ref{eq.10}) and (\ref{eq.11})
with $T_\mathrm{c}, -u, V_3$ and $V_4$, respectively.
Firstly, we try to calculate $\eta_\mathrm{max}$ by using $Q_\mathrm{h}^{\mathrm{st}}$ and 
$Q_\mathrm{c}^{\mathrm{st}}$ above. By
defining the work of one cycle $W$ as 
$W=Q_\mathrm{h}^{\mathrm{st}}+Q_\mathrm{c}^{\mathrm{st}}$, we can
calculate the efficiency $\eta=W/Q_\mathrm{h}^{\mathrm{st}}$ and the
power $P={Wu}/(2(l_3-l_1))$. The maximal power is realized at
$u=u_\mathrm{max}$ defined as a solution of $\partial P/\partial u=0$. 
Since $u_\mathrm{max}$ is given by  
\begin{eqnarray}
u_\mathrm{max}=\frac{k_\mathrm{B}(T_\mathrm{h}-T_\mathrm{c})}{4m(\frac{1}{\pi}+\frac{L}{3S})\sqrt{\frac{2\pi
 k_\mathrm{B}}{m}}(\sqrt{T_\mathrm{h}}+\sqrt{T_\mathrm{c}})},\label{eq.12}
\end{eqnarray}
$Q_\mathrm{h}^{\mathrm{st}}$ and $W$ at $u_\mathrm{max}$ are obtained as 
\begin{eqnarray}
Q_\mathrm{h}^{\mathrm{st}}(u_\mathrm{max})&&=\frac{N}{2}k_\mathrm{B}\sqrt{T_\mathrm{h}}(\sqrt{T_\mathrm{h}}+\sqrt{T_\mathrm{c}})
\ln \frac{V_2}{V_1},\\ \label{eq.13}
W(u_\mathrm{max})&&=\frac{N}{2}k_\mathrm{B}(T_\mathrm{h}-T_\mathrm{c})\ln
 \frac{V_2}{V_1}.\label{eq.14}       
\end{eqnarray}
Moreover, $\eta_\mathrm{max}\equiv \eta (u_\mathrm{max})$ is calculated
as
\begin{eqnarray}
\eta_\mathrm{max}=\frac{W(u_\mathrm{max})}{Q_\mathrm{h}^{\mathrm{st}}(u_\mathrm{max})}=1-\sqrt{\frac{T_\mathrm{c}}{T_\mathrm{h}}}.\label{eq.15}
\end{eqnarray}
This is equal to the CA efficiency Eq.~(\ref{eq.1}) though we neglect
${\cal O}(u^2)$
in the calculation of $Q_\mathrm{h}^{\mathrm{st}}$ and
$Q_\mathrm{c}^{\mathrm{st}}$. 
Therefore, we may regard that this result gives a
natural and   
microscopic foundation of the original derivation of the CA efficiency
Eq.~(\ref{eq.1}).

As seen in Fig.~\ref{fig.4}, however, the CA efficiency deviates from the MD
data and the numerically calculated line. This is 
because there exists the heat transfer other than
$Q_\mathrm{h}^{\mathrm{st}}$ and $Q_\mathrm{c}^{\mathrm{st}}$, which may
be missed in the original derivation of Eq.~(\ref{eq.1})~\cite{CA}. As seen in
Fig.~\ref{fig.2}, the initial temperatures of isothermal processes are different from
the steady values. This implies the existence of the additional heat transfer $Q_\mathrm{h}^{\mathrm{add}}$
and $Q_\mathrm{c}^{\mathrm{add}}$ during the fast relaxation to the
steady temperatures. 
Next, we repeat the similar derivation of Eq.~(\ref{eq.15})
by considering the effect of these additional heat transfers
$Q_\mathrm{h}^{\mathrm{add}}$ and $Q_\mathrm{c}^{\mathrm{add}}$.  
We define the total heat as $Q_\mathrm{h}=Q_\mathrm{h}^{\mathrm{st}}+Q_\mathrm{h}^{\mathrm{add}}$ and 
$Q_\mathrm{c}=Q_\mathrm{c}^{\mathrm{st}}+Q_\mathrm{c}^{\mathrm{add}}$. Since we assume that the relaxation to 
the steady temperature is very fast, we can approximate the additional
heat transfers as
$Q_\mathrm{h}^{\mathrm{add}}=Nk_\mathrm{B} (T_\mathrm{h}^{\mathrm{st}}-\tilde{T_\mathrm{h}})$
and $Q_\mathrm{c}^{\mathrm{add}}=Nk_\mathrm{B}
(T_\mathrm{c}^{\mathrm{st}}-\tilde{T_\mathrm{c}})$, where
$\tilde{T_\mathrm{h}}$ 
and $\tilde{T_\mathrm{c}}$ are the initial temperatures of
the isothermal processes (A) and (C), respectively. If we assume
that adiabatic processes satisfy the relations $\tilde{T_\mathrm{h}}=({V_4}/{V_1})T_\mathrm{c}^{\mathrm{st}}$ 
and $\tilde{T_\mathrm{c}}=({V_2}/{V_3})T_\mathrm{h}^{\mathrm{st}}$ which are the same as in the
quasistatic case, we can obtain
\begin{eqnarray}
&&Q_\mathrm{h}^{\mathrm{add}}=-Nk_\mathrm{B} \frac{4L\sqrt{\pi T_\mathrm{h}}}{3SA}\left(1+\sqrt{\frac{T_\mathrm{h}}{T_\mathrm{c}}}\right)u\label{eq.16} 
\end{eqnarray}
up to ${\cal O}(u)$. $Q_\mathrm{c}^{\mathrm{add}}$ is given by
changing $T_\mathrm{h}\leftrightarrow T_\mathrm{c}$ and $u \to -u$ in Eq.~(\ref{eq.16}). The work of one cycle $W$ is defined as 
$W=Q_\mathrm{h}^{\mathrm{st}}+Q_\mathrm{c}^{\mathrm{st}}+Q_\mathrm{h}^{\mathrm{add}}+Q_\mathrm{c}^{\mathrm{add}}$.
By defining $u_{\mathrm{max}}$ as a solution of $\partial P/\partial
u=0$, we can obtain
\begin{eqnarray}
u_{\mathrm{max}}&&=\frac{k_\mathrm{B} (T_\mathrm{h}-T_\mathrm{c})} 
{\sqrt{\pi}}\Biggl\{4mA\left(\frac{L}{3S}+\frac{1}{\pi}\right) \nonumber \\
&&\times
(\sqrt{T_\mathrm{h}}+\sqrt{T_\mathrm{c}})\ln \frac{V_2}{V_1}+\frac{8L
k_\mathrm{B}}{3SA\sqrt{T_\mathrm{h}}}\nonumber\\
&&\times \left(T_\mathrm{h}-T_\mathrm{c}\right)\left(1+\sqrt{\frac{T_\mathrm{h}}{T_\mathrm{c}}}\right)\Biggr\}^{-1}\ln
\frac{V_2}{V_1},\label{eq.17} \\
\eta_\mathrm{max}&&\equiv \eta(u_\mathrm{max})=\frac{\frac{N}{2}k_\mathrm{B}
(T_\mathrm{h}-T_\mathrm{c})\ln
\frac{V_2}{V_1}}{Q_\mathrm{h}^{\mathrm{st}}(u_{\mathrm{max}})+Q_\mathrm{h}^{\mathrm{add}}(u_{\mathrm{max}})}.\label{eq.18}
\end{eqnarray}
From Fig.~\ref{fig.4}, we can see that 
Eq.~(\ref{eq.18}) agrees with the MD data and the numerically calculated
result very well due to the effect of the additional heat
$Q_\mathrm{h}^{\mathrm{add}}$ and $Q_\mathrm{c}^{\mathrm{add}}$. 
To obtain the efficiency in the $T_\mathrm{c} \to T_\mathrm{h}$ limit, we set $T_\mathrm{c}=T_\mathrm{h}-\Delta
T$ $(\Delta T \ll 1)$.
Then, $\eta_\mathrm{max}$ is given by
$\eta_\mathrm{max}={\Delta T}/(2T_\mathrm{h})+{\cal O}(\Delta
T^2)$ 
which is the same as the CA efficiency up to $\Delta T$ order. 
This result explains why our $\eta_\mathrm{max}$ approaches
$\eta_\mathrm{CA}$ when $T_\mathrm{c}\to T_\mathrm{h}$. 
Very recently,
similar $\eta_\mathrm{max}$ behavior has been observed also in the other types of
the heat engines~\cite{SS,AJM}. 
In the equilibrium limit of
$T_\mathrm{c}\to T_\mathrm{h}$, the system may be regarded as being 
in the linear response regime.
Therefore, our result is consistent with the CA efficiency proved 
by using the linear response theory~\cite{VB}.
\section{Summary}
In this paper, we have studied the efficiency at the maximal power 
$\eta_\mathrm{max}$ of a finite-time Carnot cycle of a weakly
interacting gas which we can regard as a nearly ideal gas.
Our model is a natural extension of the quasistatic Carnot cycle and has
a piston moving back and forth at a constant speed $u$ in the
cylinder.
We have used this $u$ as a unique parameter to maximize the power.
Since $u$ is easily controllable, this model seems more natural than the
original Curzon-Ahlborn's model~\cite{CA}.
We have performed numerical experiments of this model by means of MD
simulations to verify the validity of the
Curzon-Ahlborn (CA) efficiency $\eta_\mathrm{CA}$ for the first time and have found that 
our $\eta_\mathrm{max}$ does not always agree with $\eta_\mathrm{CA}$,  
but approaches $\eta_\mathrm{CA}$ in the limit of 
$T_\mathrm{c} \to T_\mathrm{h}$. Our molecular kinetic 
analysis can explain the above facts theoretically by 
using only elementary arithmetic. Especially, we have revealed that the
difference between $\eta_\mathrm{CA}$ and our $\eta_\mathrm{max}$ is due
to the additional heat transfers which may be missed in the original
derivation of $\eta_\mathrm{CA}$~\cite{CA}. Though it is restricted in the equilibrium limit of
$T_\mathrm{c} \to T_\mathrm{h}$, these results strongly support the validity of
the CA efficiency from both of the experimental and
theoretical points of view.
We expect that our analysis in this paper will shed
light on the microscopic aspects of the finite-time extension 
of thermodynamics.

\acknowledgments
We thank K. Nemoto and T. Nogawa for helpful discussions.

\end{document}